\documentclass[%
 reprint,
 superscriptaddress,
 amsmath,amssymb,
 aps,
 pra
]{revtex4-2}

\usepackage{graphicx} % include figure files
\usepackage{hyperref} % add hypertext capabilities
\usepackage{dcolumn} % align table columns on decimal point
\usepackage{bm} % bold math
\usepackage[dvipsnames]{xcolor} % colors
\usepackage{braket} % Dirac notation
\usepackage{cancel} % cancel out equations
\usepackage{soul} % strikeout text
\usepackage[capitalize]{cleveref} % clever equation references
\usepackage[all]{hypcap} % make hyperlink navigate to top of figure
\usepackage{comment}
\usepackage{siunitx}
\usepackage[normalem]{ulem}

\DeclareMathOperator{\tr}{tr}

\newcommand{\HH}{\hat{H}}
\newcommand{\UU}{\hat{U}}
\newcommand{\RR}{\hat{\rho}}

\newcommand{\bb}{\hat{b}}

\begin{document}

\title{Shell-shaped Bose-Einstein condensates: Dynamics, excitations, and thermodynamics}

\author{Brendan Rhyno}
\email{rhyno@iqo.uni-hannover.de}
\affiliation{Institute of Quantum Optics, Leibniz Universität Hannover, Welfengarten 1, 30167 Hannover, Germany}
\affiliation{Department of Physics, University of Illinois at Urbana-Champaign, Urbana, Illinois 61801, USA}

\author{Kuei Sun}
\email{kuei.sun@wsu.edu}
\affiliation{Department of Physics and Astronomy, Washington State University Tri-Cities, Richland, Washington 99354, USA}

\author{Jude Bedessem}
% \email{judejb2@illinois.edu}
\affiliation{Department of Physics, University of Illinois at Urbana-Champaign, Urbana, Illinois 61801, USA}

\author{Naceur Gaaloul}
%\email{gaaloul@iqo.uni-hannover.de}
\affiliation{Institute of Quantum Optics, Leibniz Universität Hannover, Welfengarten 1, 30167 Hannover, Germany}

\author{Nathan Lundblad}
%\email{nlundbla@bates.edu}
\affiliation{Department of Physics and Astronomy, Bates College, Lewiston, Maine 04240, USA}

\author{Smitha Vishveshwara}
\email{smivish@illinois.edu}
\affiliation{Department of Physics, University of Illinois at Urbana-Champaign, Urbana, Illinois 61801, USA}

\date{\today}

\begin{abstract}
Shell-shaped Bose-Einstein condensates (BECs) represent a paradigmatic instance of quantum fluids in hollow geometries exhibiting phenomena that bridge from ultracold atomic to astrophysical realms.
In this work, we present a comprehensive survey of the dynamics, thermodynamics, and collective excitations of shell-shaped BECs, synthesizing two decades of our group's theoretical work in light of recent experimental breakthroughs.
We begin by analyzing the evolution of a BEC from filled-sphere to hollow-shell geometries, illustrating the necessity of microgravity conditions to avoid gravitational sag.
We then analyze the collective mode structure across this evolution and pinpoint a universal dip in the frequency spectra as well as mode reconfiguration due to inner-surface excitations as robust signatures of the hollowing-out transition.
Turning to vortex physics, we show that the closed-surface topology enforces vortex-antivortex configurations in shell-shaped BECs and that the natural annihilation of these pairs can be stabilized by rotation, with the critical rotation rate depending on shell thickness.
In the thermodynamic domain, we investigate the interplay between shell inflation and the BEC phase transition, where adiabatic expansions lead to condensate depletion.
This behavior motivates a study of the nonequilibrium dynamics of shell-shaped BECs; we perform such a study by constructing a time-dependent dynamic technique that can capture the evolution in both adiabatic and non-adiabatic regimes.
Finally, we review recent experimental realizations of shell-shaped BECs, including the landmark creation of ultracold shells aboard the International Space Station, and outline prospects for exploring quantum fluids in curved geometries.
\end{abstract}

\maketitle

\section{Introduction}

Quantum fluids confined in hollow geometries emerge across vastly different length scales from the mesoscopic to the astronomical~\cite{Tononi2023, Moller2020,Tononi2019,Tononi2020,Sauls1989,Pethick2017}.
In the ultracold atomic realm, shell-shaped Bose-Einstein condensates (BECs) first became relevant in optical lattice systems that trapped concentric regions of co-existent Mott insulating phases which could host condensate regions between them~\cite{DeMarco2005,Campbell2006,Sun2009}. Prospects of such structures found in microgravity or two-species mixtures of ultracold atoms have now become viable in recent experiments~\cite{Carollo2022,Jia2022,Huang2025,Guo.2022}. While condensates exhibit unique quantum features, hydrodynamic properties share plentiful commonalities with fluid bubbles found at the macroscopic level~\cite{Barcelo2011}.
In the astronomical realm, shell-shaped quantum fluids find a natural space in environments such as the interiors of neutron stars~\cite{Sauls1989,Pethick2017}, with condensates even offering prototypes for inflationary models of the early Universe~\cite{Mukhanov2005}.

Shell-shaped quantum fluids host a plethora of rich phenomena that manifest across these realms.
The topology of these hollow structures differs from their filled counterparts, leading to palpable consequences.
The presence of an inner surface, for instance, dramatically alters the nature of collective mode excitations~\cite{Lannert2007,Padavic2017,Sun2018}.
More generally, these modes can distinguish filled from hollow structures.
Quantum phenomena in hollow structures reveal unique physics, such as the geometry, stability, and dynamics of quantized vortex lines~\cite{Turner2010,Padavic2020,Caracanhas2022,Xiong.2024,White.2024,Tononi.2024wv}, and interference patterns resulting from releasing a trapped condensate bubble~\cite{Lannert2007}.
The thermodynamics of these shells proves to be complex, especially in considering condensate order as the system is tuned from a filled three-dimensional structure to an effectively two-dimensional bubble.~\cite{Rhyno2021,Tononi2022,Tononi2024}.
The notion of a geometric potential takes on importance, as does the general framing of BECs on curved manifolds~\cite{Biral.2024,Oliveira.2025,Yi.2025,Grass.2025,Costa.1981,Kaplan.1997,Zhang.2018,Bereta2019,Flachi2025},
with complementary research efforts exploring geometric effects on condensates~\cite{Kurt2024,Roy2025}, as well as additional topologies, such as M\"obius strips~\cite{Luo2025}.
For tunable shell-shaped geometries, non-equilibrium dynamics exhibits several fascinating aspects, including critical dynamics across the condensate transition and possibilities for creating the largest observed macroscopic quantum coherent bubbles~\cite{Carollo2022}.
Wavelike excitations in these systems akin to those found in planetary systems are also of interest~\cite{Saito.2023,Brito.2023,Li.2023}, as are the possibilities for studying dipolar interactions, supersolids, soliton phases for attractive interactions, or lattice models in shell systems~\cite{Adhikari.2012,Diniz.2020,Arazo.2021,Ciardi.2024,Sánchez-Baena.2024,Ghosh2024,Ciardi2025,Tononi2024_2,zhou2025xx}.

Explorations into the properties of Bose-condensed shells have seen a tremendous surge of interest since the first experimental realization of ultracold bubbles in space using the Cold Atom Lab (CAL) aboard the International Space Station (ISS)~\cite{Carollo2022, Aveline2020}.  Recent upgrades to the CAL apparatus~\cite{Williams.2024} along with future space missions on the ISS involving either the "BECCAL" device under construction~\cite{Frye2021} or other devices~\cite{Thompson.2023l5a} hold promise for near-future observations of Bose-Einstein condensation using radio-frequency (rf) dressed bubble trap geometries.
To complement perpetual free-fall experiments performed in orbital microgravity, terrestrial drop tower experiments~\cite{Zoest2010,Muntinga2013} are actively being pursued using the Einstein-Elevator at the Hannover Institute of Technology~\cite{Lotz2017,Lotz2018,Lotz2020,Lotz2023}.
In addition to achieving quantum bubbles with rf-dressed potentials, which requires gravitational sag be compensated, bubble production driven by interspecies interactions in immiscible quantum mixtures has been proposed~\cite{Wolf2022} and experimentally realized~\cite{Jia2022}.
Building upon this work, the same group recently reported observing signatures of the hollowing transition from a filled-to-hollow shell-shaped BECs~\cite{Huang2025}.

In light of these developments and more, the time is ripe for experiment and theory to work hand-in-hand in the study of shell-shaped BECs.
Theory work in our group on this topic in fact began in the early 2000's. 
It started with two of the current authors (K.S. and S.V) and colleagues in the experimental context of co-existent superfluid and Mott insulating phases in trapped optical lattice geometries~\cite{DeMarco2005,Barankov2007,Sun2009}.
On the one hand, sticking close to the excitement of new experiments at the time~\cite{Greiner2002,Campbell2006}, we characterized the spatial profiles and physical properties, such as condensate fractions and collective excitations, of superfluid shells trapped between insulating layers~\cite{Sun2009}.
On the other hand, prior to the experimental realization of condensate shells trapped in free space, we initiated characterizing their behavior as a matter of fundamental worth.
We investigated the evolution of collective modes going from the filled sphere to thin-shell limits~\cite{Padavic2017,Sun2018}, inclusive of a telltale hollowing out signature; the behavior of vortex-antivortex pairs upon rotation~\cite{Padavic2020}; and the interference patterns and accumulation of central mass density upon trap release and time-of-flight~\cite{Lannert2007,Sun2009}.
Over a decade back, one of the current authors (N.L) and colleagues began pioneering the microgravity experiments mentioned above, aboard the ISS~\cite{Aveline2018,Aveline2020,Carollo2022}. In collaboration, theoretical work~\cite{Rhyno2021} performed by current authors (N.L., B.R. and S.V) and colleagues, described thermodynamics features of the ultracold quantum gas shells (or bubbles) created in the experiments~\cite{Carollo2022}. 

These theoretical works dating back from over two decades to our current studies have now been realized and described in multiple contexts.
To name a few, the spectral signature encoded in the collective modes as the shell transitions from a filled-to-hollow sphere has received extended analysis in experiments with dual-species mixtures~\cite{Jia2022,Huang2025}.
A recent perspective article~\cite{Dubessy2025} on creating shell-shaped condensates offers an overview of various physical features of these shells, briefly including our descriptions of the density profiles, vortex-antivortex structure and thermodynamics.
As we neither have all our treatments combined as a single treatise nor have we put forward our most recent findings, in this timely moment, we present here a comprehensive survey of our past studies and latest new results on the dynamic evolution of condensate shells.

In Sec.~\ref{sec:Equilibrium properties}, we study the equilibrium density profiles of BECs in trapping potentials capable of producing shell-shaped geometries.
We show how a bubble-trap potential enables smooth evolution from filled-sphere to hollow-shell geometries and how a microgravity environment is essential for maintaining the topology of a shell in rf-dressed shell-shaped BECs.
In Sec.~\ref{sec:Collective modes}, we study the collective motion of a BEC, which manifests as oscillations of the condensate density around equilibrium and reflects the underlying shell-shaped geometries in its mode structure.
We find features in the collective-mode spectrum of filled-sphere and hollow-shell geometries as well a robust signature of the hollowing-out transition.
In Sec.~\ref{sec:Vortex physics}, we touch upon vortex physics in shell-shaped BECs, where the hollow topology imposes unique constraints that require vortex-antivortex pairs to satisfy the zero net circulation. Such a pair is driven toward annihilation by an attractive interaction, while it can be stabilized in a rotating condensate. The critical rotation speed required for stabilization increases with the shell thickness, suggesting an experimental method to determine the thickness.

In Sec.~\ref{sec:Thermodynamics}, we explore the thermodynamic properties of low-density BEC bubbles.
We compute the critical temperature of the system over the range of possible geometries from filled-to-hollow shells, and also determine the evolution of the temperature and condensate fraction during isentropic expansions in which one observes a loss of space density.
In Sec.~\ref{sec:Nonequilibrium_dynamics}, we explore the nonequilibrium properties of low-density BEC bubbles.
Starting from a finite temperature filled sphere BEC, we dynamically inflate the bubble to the thin-shell regime and monitor the instantaneous condensate and excited state fractions.
Depending on the rate at which the quench is performed, we show examples where the system remains close to its instantaneous ground state and also observe decaying oscillations in the condensate fraction below equilibrium predictions. 
Finally, in Sec.~\ref{sec:Experimental_realizations}, we discuss the recent experimental progress on producing quantum bubbles and the future outlook.

\section{Equilibrium properties}\label{sec:Equilibrium properties}

\begin{figure}[t]
  \centering
  \includegraphics[width=8.6cm]{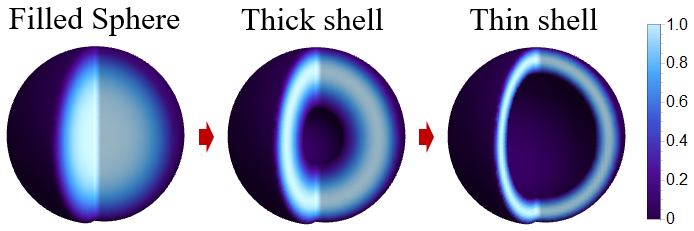}
  \caption{
    (Color online)
    Schematic density profiles $n_{\textrm{eq}}(\mathbf{r})$ of a Bose-Einstein condensate evolving from filled-sphere to hollow-shell geometries.
       (Adapted from Ref.~\cite{Sun2018}. Copyright (2018) by the American Physical Society.)
  }
  \label{fig:equilibrium_fig_1}
\end{figure}

Shell-shaped Bose-Einstein condensates (BECs) represent a unique class of quantum fluids that differ significantly from their filled-sphere counterparts. The primary distinction lies in the presence of an internal surface in shell geometries, which fundamentally alters the equilibrium and dynamical properties of the condensate. As the trapping potential is tuned, a condensate can transition from a filled sphere to a hollow shell, as shown in Fig.~\ref{fig:equilibrium_fig_1}, with a hollowing-out transition where the central density depletes to zero and the inner surface develops.

To explore BECs, we start with the Hamiltonian for a weakly interacting bosonic gas,
\begin{align}
    \label{eq:Hamiltonian}
    \hat H = \int d\mathbf{r}
    \bigg[& \hat \psi^\dagger(\mathbf{r}) \left( -\frac{\hbar^2}{2m}\nabla^2 + V(\mathbf{r}) \right) \hat \psi(\mathbf{r})
    \nonumber\\
    &+
    \frac{U_0}{2}  \hat\psi^\dag(\mathbf{r}) \hat\psi^\dag(\mathbf{r}) \hat\psi(\mathbf{r}) \hat\psi(\mathbf{r}) \bigg]
    ,
\end{align}
where $\hat \psi^\dagger(\mathbf{r})$ ($\hat \psi(\mathbf{r})$) is the creation (annihilation) operator for a bosonic atom at position $\mathbf{r}$, $m$ is the particle mass, $V(\mathbf{r})$ is the trapping potential, and
$U_0=4 \pi \hbar^2 a_s /m$ is the interaction strength
(proportional to the two-body scattering length $a_s$)
\cite{Pethick2008}. With a mean-field treatment, the equilibrium properties of BECs are characterized by wavefunctions $\psi(\mathbf{r})$ obeying the time-independent Gross-Pitaevskii (GP) equation,
\begin{eqnarray}\label{eq:GPeqn}
\left[ { - \frac{{{\hbar
^2}}}{{2m}}{\nabla ^2} + V({\mathbf{r}}) + U_0{{\left| {\psi
({\mathbf{r}})} \right|}^2}} \right]\psi ({\mathbf{r}})=\mu \psi(\mathbf{r}),
\end{eqnarray}
where $\mu$ is the chemical potential. The equilibrium condensate density, given by $n_{\textrm{eq}}(\mathbf{r}) = |\psi ({\bf{r}})|^2$, defines its geometry and is shaped by the interplay between confining potentials $V({\bf{r}})$, interatomic interactions $U_0$, and external influences such as gravity. It also plays a key role in determining the collective modes of the condensate. 

As the simplest of scenarios, a filled-sphere condensate can be realized with a spherical harmonic trap $V(r)=\frac{1}{2}m\omega_0^2 r^2$, where $\omega_0$ is the trapping frequency, and $r$ is the radial distance from the spherical center. This trap has its potential minimum at the spherical center, where particles naturally concentrate to form a filled density profile. Building on possibilities offered by the harmonic trap, a shell-shaped condensate can be realized by radially shifting the potential minimum to a nonzero radial distance $r_0$, obtaining a trapping potential with a characteristic frequency $\omega_{\text{sh}}$ as
\begin{equation} \label{eq:off_center_harm}
    V_{\text{sh}}(r)=\frac{1}{2}m\omega_{\text{sh}}^2(r-r_0)^2.
\end{equation}
A sufficiently dilute condensate in this trap exhibits a thin shell geometry with its radius $\approx r_0$.
While the radially shifted potential is a good approximation for studying several equilibrium and dynamic properties of thin shell-shaped condensates, it is unphysical close to $r=0$ where the slope of the potential becomes discontinuous.
Thus, one cannot use it to study the evolution between filled-sphere and hollow-shell geometries by varying $r_0$.

\begin{figure}[t]
  \centering
  \includegraphics[width=8.6cm]{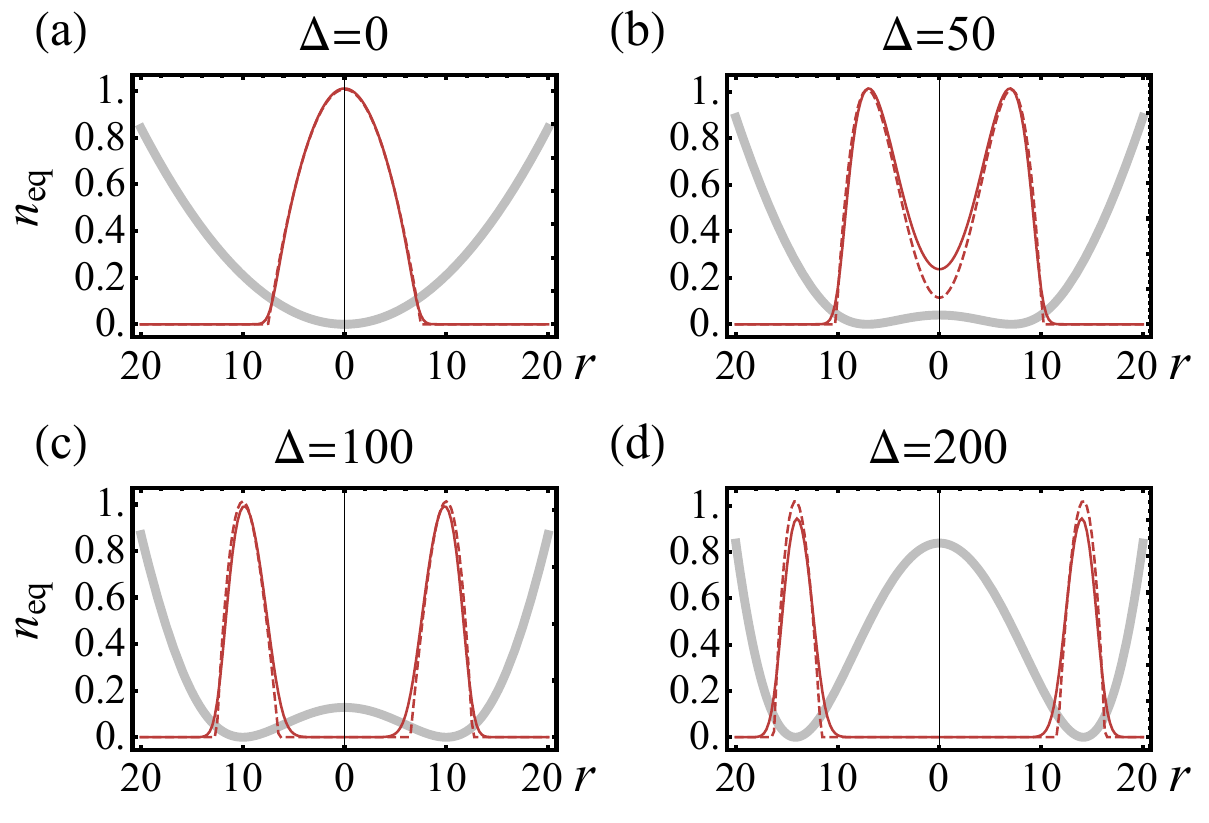}
  \caption{
    (Color online) (a--d) Equilibrium density profiles $n_{\textrm{eq}}(r)$ (thin curves; with the maximum value set to 1) of a Bose-Einstein condensate and the corresponding bubble-trap potential (thick curves; in arbitrary units) for various detuning $\Delta$, showing the evolution from the filled-sphere geometry at
    $\Delta=0$ to a hollow thin-shell one at a large $\Delta$. The solid profiles show the numerical solutions of the GP equation, while the dashed ones are based on the Thomas-Fermi approximation.
  }
  \label{fig:equilibrium_fig_2}
\end{figure}

As is our focus here, a more realistic trapping potential producing shell-shaped condensate and a smooth sphere-to-shell evolution is a ``bubble trap"~\cite{Zobay2001,Colombe2004,Merloti2013}. Cold atomic experiments can achieve such a trapping potential by employing time-dependent rf-dressed
adiabatic potentials within a conventional magnetic trapping geometry. The trapping potential has the form 
\begin{equation}\label{eq:bubble}
V_{\textrm{bubble}}(r)=m\omega_0^2
\sqrt{(r^2-\Delta)^2/4+\Omega_b^2},
\end{equation}
where $\Delta$ represent the effective detuning between the applied rf field and the energy states used to prepare the condensate and $\Omega_b$ is proportional to the Rabi coupling between these states.
The unit of length is chosen to be the oscillator length $S_l \equiv \sqrt{\hbar/2m\omega_0}$.
The bubble trap has its potential minimum at $r=\sqrt{\Delta}$, which enables a smooth evolution between the filled sphere and hollow shell geometries by tuning $\Delta$. At $\Delta=0$, the bubble-trap potential reduces to a spherically harmonic trap characterized by frequency $\omega_0$, producing a filled-sphere condensate. For large $\Delta$, the trap is approximated near its minimum by the radially shifted harmonic potential of Eq.~(\ref{eq:off_center_harm}) with  frequency $\omega_{\text{sh}}=\omega_0 \sqrt{\Delta/\Omega_b}$. To illustrate salient features of condensate shells,  we employ a single tuning variable by setting $\Omega_b = \Delta$ for the following calculations and discussions involving the bubble trap. Figure \ref{fig:equilibrium_fig_2} shows the equilibrium density profiles $n_{\textrm{eq}}(r)$ of a condensate in the bubble trap for various values of the detuning parameter $\Delta$. By slowly changing $\Delta$, one can realize a continuous deformation between the two limiting geometries and a topological transition from filled to hollow sphere at a critical value.

In addition to numerically calculating the equilibrium density profile from the GP equation of Eq.~(\ref{eq:GPeqn}), one can apply the Thomas-Fermi approximation to find a well approximated solution in the limit of strong interactions. The approximation neglects the  kinetic energy term as it is small compared to the interaction energy, thereby obtaining the equilibrium density profile
\begin{eqnarray}\label{eq:neq}
n_{\textrm{eq}}(\mathbf{r})=\frac{V(\mathbf{R})-V(\mathbf{r})}{U_0}.
\end{eqnarray}
 The Thomas-Fermi density profile exhibits a clear boundary $n_{\textrm{eq}}(\mathbf{R})=0$, determined by the trap geometry and the particle number $N = \int {{n_{{\textrm{eq}}}}({\bf{r}})d\bf{r}}$. For a shell-shaped condensate in the bubble trap, the Thomas-Fermi density profile exhibits an outer radius $R$ and an inner radius $R_{\textrm{in}}=\sqrt{2\Delta-R^2}$. The hollowing out transition therefore occurs at $\Delta = R^2/2$.

In Fig.~\ref{fig:equilibrium_fig_2}, we show the equilibrium density profiles in the bubble trap obtained from the numerical calculation of the GP equation (solid curves) and the Thomas-Fermi approximation (dashed curves) for a condensate of $\sim 10^4$ $^{87}$Rb atoms in a bubble of bare frequency $\omega_0/2\pi=$ 10--100 Hz.
The exact numerical calculations show a more realistic, smooth decrease in density at the edges of the condensate, but the Thomas-Fermi profiles match the former well in the range of realistic parameters shown.
The Thomas-Fermi approximation thus provides a convenient and effective way to analyze the condensate's static and dynamic properties such as the influence of gravity and the structure of collective modes.

Under terrestrial conditions, gravity introduces a non-negligible perturbation on rf-dressed shell-shaped condensates as it distorts the shell structure via gravitational sag. 
The gravitational force pulls the condensate downward, resulting in mass accumulation at the lower vertical points in the system and depletion around the highest. To model this sag for shell-shaped condensates, we introduce a gravitational term $-gr\cos{\theta}$ to the potential. In the weak gravity and thin-shell limits, there is a critical gravitational acceleration
\begin{equation}
    g_c = \frac{(15U\omega N)^{2/3}}{(128m\pi)^{2/3}r_0^{7/3}S_l^{1/3}},
\end{equation}
above which a shell with $N$ particles will collapse as the local density at its highest point depletes to zero~\cite{Sun2018}. For the specific case of the bubble trap potential, the single-particle oscillation frequency is given by $\omega=\omega_0\sqrt{\Delta/\Omega_b}$, and the unperturbed radius is $r_0=\sqrt{\Delta}$.

\begin{figure}[t]
  \centering
  \includegraphics[width=8.7cm]{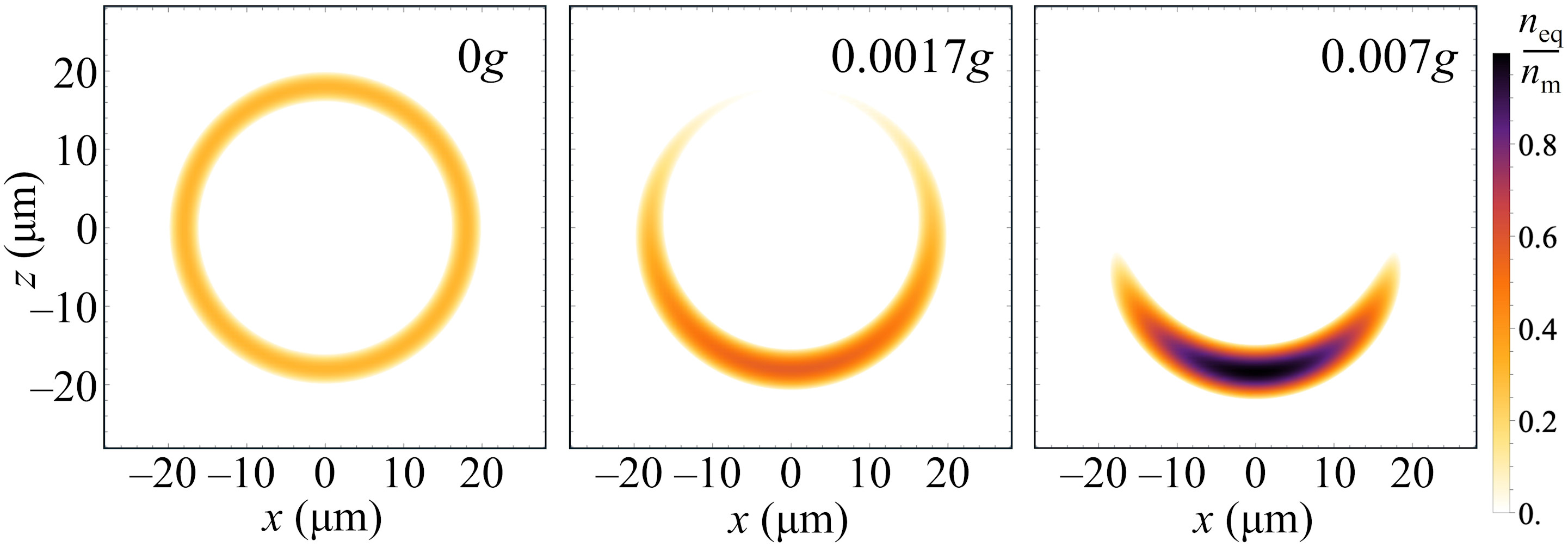}
  \caption{
    (Color online) Thomas-Fermi density profiles in the $x$-$z$ plane for condensates confined by a bubble trap without gravity (left) and under the influence of gravitational fields $0.0017g$ (middle) and $0.007g$ (right), where $g$ is the gravitational acceleration on Earth (in the $-z$ direction). These profiles are generated for $10^5$ $^{87}$Rb atoms in a bubble trap approximated by Eq.~(\ref{eq:off_center_harm}) with $\omega_{\rm{sh}}=403$ Hz, forming a condensate shell with outer radius $20$ $\mu$m and thickness $4$ $\mu$m in the absence of gravity. The colors in the bar graph represent density normalized by $n_m=2.96 \times 10^{13}/\textrm{cm}^{3}$. As the strength of the gravitational field increases, we observe a density depletion at the top of the condensate shell and a density maximum at its bottom.
  }
  \label{fig:equilibrium_fig_3}
\end{figure}

In Fig.~\ref{fig:equilibrium_fig_3}, we show the equilibrium density profile of a realistic shell-shaped $^{87}$Rb condensate for various gravitational accelerations.
We see that the shell shape is not maintained even at $10^{-3}g$.
Microgravity platforms such as CAL aboard the ISS and the Einstein-Elevator are therefore essential for realizing and maintaining spherically symmetric hollow BECs.

In summary, we have examined the equilibrium density profile of shell-shaped BECs as a result of the interplay between the trapping potential, interatomic interactions, and external influences such as gravity. We analyzed a bubble-trap potential, realized experimentally using rf-dressed magnetic fields, that enables filled sphere and thin-shell geometries of a condensate as well as a smooth evolution between them. We showed how the hollowing-out transition occurs as the bubble-trap parameters vary. Finally, we found how gravity distorts the shell symmetry by sagging the condensate, with collapse occurring beyond a critical gravitational strength. This highlights the necessity of microgravity environments to sustain symmetric shell-shaped BECs.

\section{Collective modes}\label{sec:Collective modes}

Probing dynamics offers rich insights into the properties of Bose-Einstein condensates. The elementary excitations of trapped BECs behave as a collective motion of particles across the system and are commonly accessed by experiment.
In this section, we study the collective motion of shell-shaped BECs and show how the collective-mode frequencies reflect the geometry of the system. Our analysis tracks the evolution of a BEC from filled-sphere to hollow-shell geometries through the behavior of the collective-mode spectrum, where distinct signatures point out the hollowing-out transition.

The dynamics of a BEC is described by the time-dependent GP equation (same form as Eq.~(\ref{eq:GPeqn}) with $\mu$ replaced by $i\hbar\partial_t$) for its wavefunction $\psi(\mathbf{r},t)$. Equivalently, hydrodynamic equations can be obtained in terms of its density $n(\mathbf{r},t)=|\psi(\mathbf{r},t)|^2$ and phase $S(\mathbf{r},t)=\textrm{Arg}[\psi(\mathbf{r},t)]$ (which is associated with the flow velocity $\mathbf{v}=(\hbar/m)\nabla S$). To find the collective-mode frequencies and the corresponding density deviations, we consider an oscillation density profile of the form 
\begin{eqnarray}\label{eq:time_dependent_density}
 n(\mathbf{r},t) = n_{\textrm{eq}}(\mathbf{r})+\delta n(\mathbf{r}) e^{i \omega t}.
\end{eqnarray}
Substituting Eq.~(\ref{eq:time_dependent_density}) into the hydrodynamic equations and linearizing the equations by treating the density deviation $\delta n$ and velocity $\mathbf{v}$ as small quantities (see the treatment in Ref.~\cite{Pethick2008}), we obtain an eigenvalue problem for the collective modes,
\begin{eqnarray}
\label{eq:eigen_problem_1}
- \frac{m}{U_0} \omega^2 \delta n(\mathbf{r})
    = \nabla n_{\textrm{eq}} \cdot  \nabla \delta n + n_{\textrm{eq}}\nabla^2 \delta n.
\end{eqnarray}
Thus, the eigen solutions are determined by the equilibrium density profile $n_{\textrm{eq}}(\mathbf{r})$ and its spatial derivative $\nabla n_{\textrm{eq}}(\mathbf{r})$. The oscillation spectrum therefore hinges on the specific geometry of the condensate. 

For a spherically trapped BEC, the solution to Eq.~(\ref{eq:eigen_problem_1}) is quantized with a set of radial quantum numbers $\nu$ and angular ones $(l,m_l)$.
The density deviation profile is decoupled in the radial and angular directions as $\delta n_{\nu l m_l} (\mathbf{r}) = D_{\nu l}(r) \, Y^{m_l}_l(\theta,\phi)$. Here the angular component $Y^{m_l}_l(\theta,\phi)$ are spherical harmonics, and the radial one $D_{\nu l}(r)$ obeys
\begin{align}\label{eq:eigen_problem_2}
    \frac{m}{U_0} \omega_{\nu l}^2 r^2 D_{\nu l}
    =
    - \partial_r
    \left( r^2 n_{\textrm{eq}} \partial_r D_{\nu l} \right)
    + l(l+1) n_{\textrm{eq}} D_{\nu l}.
\end{align}
Using the bubble-trap potential of Eq.~(\ref{eq:bubble}) with $\Omega_b=\Delta$ and the corresponding Thomas-Fermi equilibrium profile of Eq.~(\ref{eq:neq}), we are equipped to solve Eq.~(\ref{eq:eigen_problem_2}) for the collective modes of a condensate as it evolves from a filled sphere to hollow shell geometry.

\begin{figure}[t]
  \centering
  \includegraphics[width=7cm]{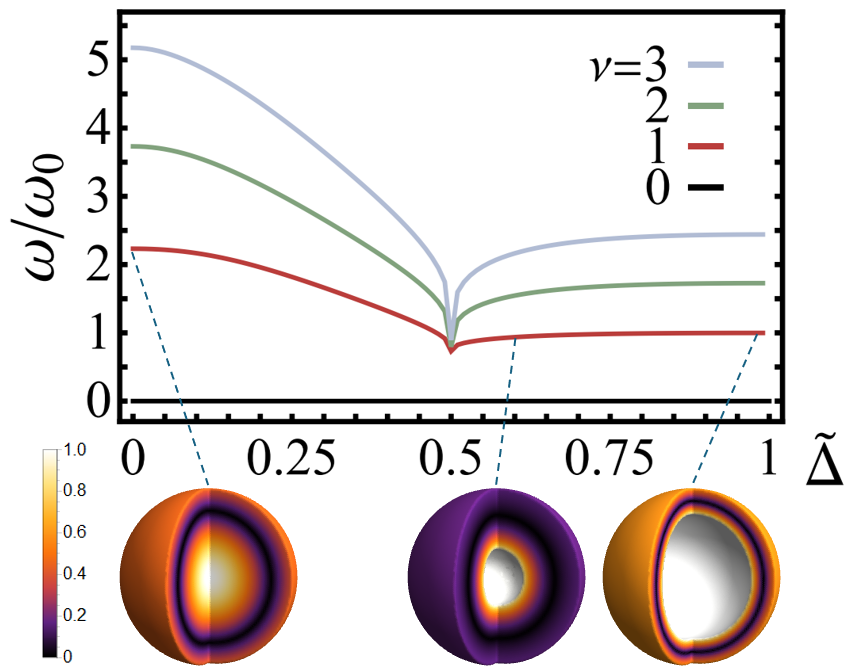}
  \caption{
    (Color online) Oscillation frequencies $\omega$ of the three lowest-lying ($\nu=1,2,3$) spherically symmetric ($l=0$) collective modes vs the normalized bubble-trap detuning $\tilde \Delta$ ($\equiv \Delta/R^2$).
    The zero frequency curve ($\nu=0$ mode) is presented for comparison. The condensate evolves from a filled sphere ($\tilde \Delta=0$) toward a hollow thin shell ($\tilde \Delta \to 1$), through a hollowing transition at $\tilde \Delta=0.5$, where the frequency curves exhibit a dip due to the appearance of a sharp new boundary in the density profile. Insets: schematic density oscillation profiles $\delta n(r)$ of the corresponding collective modes of the filled sphere and hollow shell condensates. (Adapted from Ref.~\cite{Sun2018}. Copyright (2018) by the American Physical Society.)
  }
  \label{fig:collective_fig_1}
\end{figure}

For the standard fully filled sphere geometry corresponding to the bubble-trap detuning $\Delta=0$, the collective-mode frequencies have the form $\omega_{\nu l}/\omega_0=\sqrt{l+3 \nu+2\nu l+2 \nu^2}$~\cite{Stringari1996}.
For $\Delta \neq 0$, the eigenproblems can be solved numerically.
In Fig.~\ref{fig:collective_fig_1}, we show the collective-mode frequencies of the first three modes, $\nu=1$, 2, 3 for $l=0$, as a function of the normalized bubble-trap detuning $\tilde \Delta \equiv \Delta/R^2$, which evolves from 0 (filled sphere) to 1 (infinitesimally thin shell) with the hollowing-out transition at $\tilde \Delta = 0.5$. These $l=0$ modes correspond to spherically symmetric oscillations, with density deviations $\delta n$ only in the radial direction (see the insets), and are called ``breathing'' modes. The different $\nu$ modes have different $\nu$ oscillation nodes (where $\delta n=0$) in the radial direction. We find that for $0 < \tilde \Delta < 0.5$, where the sphere starts depleting from its center but remains filled, the mode frequencies decrease with $\tilde \Delta$. For $\tilde \Delta > 0.5$, where the inner boundary appears and the system becomes shell-shaped, the mode frequencies increase with $\tilde \Delta$. A sharp dip is developed for all the modes at $\tilde \Delta = 0.5$. This marks a clear signature of the hollowing-out transition. 

To understand the behavior of the frequency curves, we carefully consider the boundary conditions of the eigenproblem. We have seen in Eq.~(\ref{eq:eigen_problem_1}) that the two factors $n_{\textrm{eq}}(\mathbf{r})$ and $\nabla n_{\textrm{eq}}(\mathbf{r})$ determine the solutions to the eigenproblem. Before the hollowing-out transition, we have $n_{\textrm{eq}} \neq 0$ and $\nabla n_{\textrm{eq}} = 0$ at the center of the filled condensate. After the transition, the condition suddenly changes to $n_{\textrm{eq}} = 0$ and $\nabla n_{\textrm{eq}} \neq 0$ at the inner boundary of the shell. The emergence of the inner boundary and the sudden change in the boundary of the equilibrium density profile cause the change in the behavior of the frequency curves. At the transition, both $n_{\textrm{eq}} = \nabla n_{\textrm{eq}} = 0$, and we therefore need to consider the higher-order effects beyond the Thomas-Fermi approximation. In general, the density profile does not sharply deplete out but develops a tail that gradually decays over space. The decay rate, which depends on the competition between the kinetic energy and the trapping potential, largely affects the frequency spectrum around the transition point. Our calculations show that the sharpness of the frequency dip varies with trapping potentials producing different density-profile decay rates, but the dip structure in the curve is robust at the hollowing-out transition regardless of the potential details. Such a dip feature in the collective-mode frequency spectrum marks a universal and unequivocal signature of the hollowing-out transition.

\begin{figure}[t]
  \centering
  \includegraphics[width=7cm]{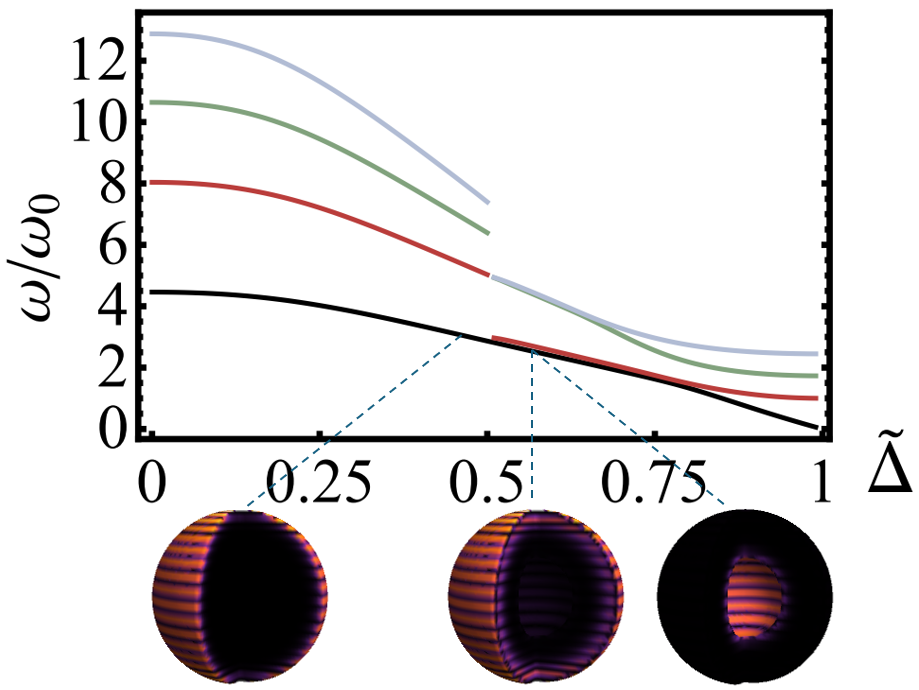}
  \caption{
    (Color online) Oscillation frequencies $\omega$ of the four lowest-lying ($\nu=0,1,2,3$) large-angular-momentum ($l=20$) collective modes vs the bubble-trap detuning $\tilde \Delta$, with the same convention as in Fig.~\ref{fig:collective_fig_1}.  Insets: schematic density  oscillation profiles $\delta n(\mathbf{r})$ showing one outer-surface mode on the filled condensate and two inner-surface and outer-surface modes on the shell-shaped condensate.
    (Adapted from Ref.~\cite{Sun2018}. Copyright (2018) by the American Physical Society.)
    }
  \label{fig:collective_fig_2}
\end{figure}

We see another interesting feature in the large-$l$ modes, which directly indicates the presence of the inner boundary in shell-shaped BECs. In Fig.~\ref{fig:collective_fig_2}, we plot the first four $l=20$ modes vs $\tilde \Delta$. We see that in the filled region or $\tilde \Delta < 0.5$, the curves behave similarly to the $l=0$ modes in Fig.~\ref{fig:collective_fig_1} (except the $\nu=0$ mode becomes nonzero due to the angular motion). When extending to the shell-shaped region or $\tilde \Delta > 0.5$, each curve splits into two: the $\nu=0$ curve in the filled region splits into the $\nu=0,1$ curves in the shell-shaped region, the $\nu=1$ curve splits into the $\nu=2,3$ curves, and in general, the $\nu=n$ curve splits into the $\nu=2n,2n+1$ curves. To understand such splitting, we look at the density deviation profiles of the collective modes. As shown by the insets in Fig.~\ref{fig:collective_fig_2}, the large-$l$ modes are ripples on the surfaces of the condensate. A filled condensate has only the outer surface to accommodate the large-$l$ modes. In a shell-shaped condensate, the outer-surface mode experiences a smooth evolution from the filled case, while a new mode appears on the emerging inner surface. This new branch of the inner surface mode directly distinguishes the shall-shaped geometry from the filled one.

In the thin-shell limit, we can solve for the density fluctuations analytically. In this limit, any spherically symmetric trapping potential can be approximated by the shifted harmonic potential $V_{\text{sh}}(\mathbf{r})$ in Eq.~(\ref{eq:off_center_harm}). We substitute the Thomas-Fermi approximation corresponding to $V_{\text{sh}}(\mathbf{r})$ into Eq.~\eqref{eq:eigen_problem_1}. The resulting expression recovers the Legendre equation in the thin-shell limit having solutions given by $\delta n(\mathbf{r})=\sqrt{\frac{2\nu + 1}{2}}P_{\nu}\left(\frac{r-r_0}{\delta}\right) Y_l^{m_l}(\theta, \phi)$
and associated frequencies
\begin{equation}
    \omega_{\nu,l}=\omega_{\text{sh}}\sqrt{\nu(\nu+1)/2}.
\end{equation}

In the presence of a gravitational field, we would expect the above collective modes to mix because of the $-mgz$ modification to the effective trapping potential. Under the influence of gravity,  Eq. \eqref{eq:eigen_problem_1} takes the modified form
\begin{equation}
\begin{split}
     \omega^2\delta n (\mathbf{r}) &= \frac{1}{mS_l^2}\frac{\partial V_{\text{sh}}(\mathbf{r})}{\partial r}\frac{\partial \delta n(\mathbf{r})}{\partial r} - \frac{\mu-V_{\text{sh}}}{mS_l^2}\nabla^2\delta n(\mathbf{r} ) 
     \\
     &+\frac{g}{S_l}\left[\cos\theta \frac{\partial }{\partial r}-\frac{1}{r}\sin\theta\frac{\partial}{\partial\theta}-r\cos\theta \hspace{2mm}\nabla^2\right]\delta n(\mathbf{r})
     .
\end{split} 
\end{equation}
The perturbations to the frequency spectrum  $\langle \delta n (\mathbf{r})_{\nu, m_l}^l| V_g(r, \theta)|\delta n (\mathbf{r})_{\nu', m_l'}^{l'}\rangle$ vanish to first order in the thickness of the condensate shell, $\delta$, unless $\{\nu, l, m_l\}=\{\nu', l'\pm1, m_l'\}$.  A weak gravitational perturbation thus has the effect of mixing adjacent angular momentum modes.

In summary, we studied how the collective modes of a BEC are determined by its equilibrium density profile and can be used to probe the formation of shell-shaped BECs. We analyzed the spectrum of radial and angular oscillations for a BEC in a bubble trap evolving from filled-sphere to thin-shell geometries. As the system transitions to a hollow shell, the appearance of an inner boundary leads to striking spectral signatures: the radial breathing modes develop a frequency dip at the hollowing transition, while the surface modes split into inner- and outer-surface branches. These robust features provide unambiguous markers of the transition from filled-to-hollow states. Finally, we found that the presence of gravity mixes modes of neighboring angular momentum, illustrating how realistic perturbations affect the collective dynamics.

\section{Vortex physics}\label{sec:Vortex physics}

\begin{figure}[t]
  \centering
  \includegraphics[width=8cm]{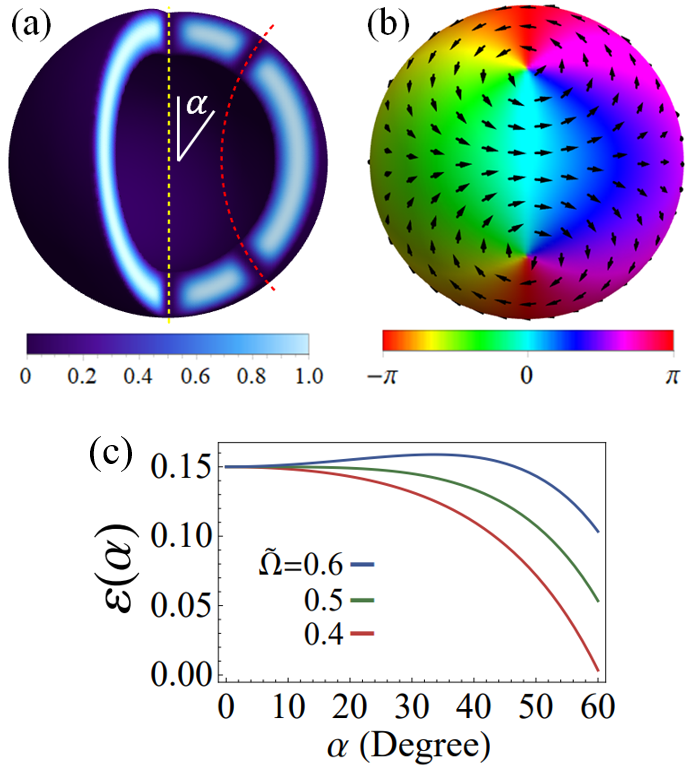}
  \caption{
    (Color online) (a) Schematic density profile of a shell-shaped Bose-Einstein condensate hosting a vortex-antivortex pair at the poles and another such pair with the vortex at polar angle $\theta = \alpha$ and antivortex at $\pi - \alpha$, rotating about the dashed lines. (b) The superfluid flow field showing the vortex-antivortex pair at $\theta = \alpha$. The colors represent the phase of the condensate wave function. (c) Dimensionless superfluid flow energy $\varepsilon (\alpha)$ of a rotating condensate shell for dimensionless angular velocities $\tilde \Omega$ above, equal to, and below the critical value $\tilde \Omega_c = 0.5$, showing a local energy minimum develops for the polar vortex-antivortex pair at a sufficient rotation rate.
    (Reused from Ref.~\cite{Padavic2020}. Copyright (2020) by the American Physical Society.)
  }
  \label{fig:vortex_fig_1}
\end{figure}

Quantized vortices are a hallmark of superfluids in response to rotations and are common probes of superfluidity in BEC experiments. The single-valuedness of the condensate wavefunction imposes a quantization condition that the circulation number of the superfluid velocity field $\ell_v = (m/\hbar) \oint \mathbf{v} \cdot d\bm{l}$ must be integer, where $\mathbf{v}=(\hbar/m)\nabla S$, is the phase gradient of the wavefunction. In the case of nonzero circulation, the GP equation captures the existence of topological defects where the wavefunction vanishes, forming vortex cores.  In filled BECs, the topological defect forms a vortex line throughout the system. As a filled BEC hollows out, the rotational flow field still behaves as if it rotates about an invisible line, as shown in Fig.~\ref{fig:vortex_fig_1}(a). Observed along the BEC's outer surface,  the defects form vortex-antivortex pairs on the shell-shaped geometry.

In the thin-shell limit, the system becomes a two-dimensional (2D) closed surface, where its $S^2$ topology poses a strict constraint on the allowed vortex configurations. To elaborate,  consider a condensate shell that accommodates multiple vortices. Any closed loop on the shell that encloses an area $A$ also encloses the complementary area $A'$ in the opposite direction. The velocity loop integral $\oint \mathbf{v} \cdot d\bm{l}$ shows that the total circulation number in $A$ is the negative of the total circulation number in $A'$. Therefore, the total circulation number of the shell condensate is always zero. Given $n_v$ vortices and the $i$th vortex having vorticity (circulation number) $\ell_i$, we write down the constraint as
 \begin{eqnarray}\label{eq:circulation_constraint}
\sum\nolimits_{i=1}^{n_v} \ell_i =0.
 \end{eqnarray}
With this constraint, the simplest vortex configuration on a shell condensate is $n_v=2$ and $(\ell_1,\ell_2)=(1,-1)$, corresponding to a vortex-antivortex pair as shown in Fig.~\ref{fig:vortex_fig_1}(b). Note that the constraint applies not only to a 2D shell but also a thick shell as long as each vortex pierces through the shell from the outer to inner surfaces.

Two vortices interact as their flow fields overlap each other. If the energy of the resultant flow field decreases as the two vortices approach each other, then an attractive interaction exists between them; conversely, if the energy increases, then a repulsive interaction exists. For two point-like vortices in a 2D superfluid with particle density $\rho_\textrm{2D}$, the effective interaction energy can be evaluated as $\frac{\hbar^2\rho_\textrm{2D}}{2m^2} \int |\nabla S|^2 dA$~\cite{Turner2010}. Using this, we calculate the interaction energy for a vortex-antivortex pair residing at $(\theta,\phi)=(\alpha,0)$ and $(\pi-\alpha,0)$ on a shell condensate, as illustrated in Fig.~\ref{fig:vortex_fig_1}(a), and obtain
\begin{eqnarray}\label{eq:vortex_interaction_on_shell}
E_{\textrm{v-av}} (\alpha) = \frac{\pi\hbar^2\rho_\textrm{2D}}{m^2} \ln(\cos \alpha).
\end{eqnarray}
The energy monotonically decreases with an increasing $\alpha$, which means an decreasing angular separation ($\pi-2\alpha$) between the vortex and the antivortex, thus showing an attractive interaction.
In the presence of energy and angular-momentum relaxing mechanisms, a vortex and antivortex on a shell condensate tend to merge and annihilate each other, i.e. vortex-antivortex pairs are unstable.
We note that due to the spherical symmetry, Eq.~(\ref{eq:vortex_interaction_on_shell}) works for the vortex-antivortex separation in any polar and azimuthal directions (with the proper coordinate change for $\alpha$). However, the equation holds only if the separation is large compared with the superfluid coherence length. When the vortex and antivortex nearly merge, or $\alpha \to \pi/2$, additional corrections are needed to avoid divergences in energy.

In experiments, vortices are usually produced and studied in a rotating condensate, where the formation of vortices is energetically favorable if the vortex angular momentum is aligned with the condensate's rotational velocity.
In a shell condensate rotating about the $z$ axis, the energy of a vortex-antivortex pair is lowered the most if the vortex stays at the north pole and its partner at the south pole, i.e. the $\alpha=0$ case in Fig.~\ref{fig:vortex_fig_1}(a). Since $\alpha=0$ corresponds to the energy maximum (an unstable state) in the non-rotating system, we expect that rotation can stabilize the vortex-antivortex pair in shell condensates. To verify this argument, we calculate the vortex energy in a rotating shell by adding the rotation-induced energy $-\Omega_\textrm{rot} \langle L_z \rangle $ to Eq.~(\ref{eq:vortex_interaction_on_shell}), where $\Omega_\textrm{rot}$ is the angular velocity of rotation assumed in the $+z$ direction and $L_z$ is the $z$-component angular momentum operator. We express the energy per particle in a dimensionless form as
\begin{eqnarray}\label{eq:vortex_energy_rotation}
\varepsilon (\alpha)= \frac{E_{\textrm{v-av}}-\Omega_\textrm{rot} \langle L_z \rangle}{N \hbar^2 / 2mR^2} = \frac{1}{2}\ln(\cos\alpha)-\tilde{\Omega}\cos\alpha.
\end{eqnarray}
Here, $N$ is the total number of particles, $R$ is the shell radius, and $\tilde{\Omega}=\frac{2mR^2}{\hbar} \Omega_\textrm{rot}$ represents a dimensionless angular velocity.
In Eq.~(\ref{eq:vortex_energy_rotation}), we see that the rotation-induced energy $-\tilde{\Omega}\cos\alpha$, which monotonically increases with angle $\alpha$, competes with the non-rotating energy $\frac{1}{2}\ln\cos\alpha$. As a result, there exists a critical rotation $\tilde{\Omega}_c=0.5$, such that the energy remains monotonically decreasing with $\alpha$ in $\tilde\Omega\le\tilde\Omega_c$ but exhibits a local minimum at $\alpha=0$ in $\tilde\Omega > \tilde\Omega_c$. In Fig.~\ref{fig:vortex_fig_1}(c), we plot the energy curves around the critical value to show such a transition. The newly emerging energy minimum for $\tilde\Omega > \tilde\Omega_c$ represents a metastable state, showing that a sufficiently large rotation can stabilize a vortex-antivortex pair at the poles against self-annihilation.

Having analyzed vortex physics in a 2D shell geometry, we now turn to study a more realistic three-dimensional (3D) shell condensate with thickness $\delta$, where the vortex core forms a string threading through the shell rather than a point-like defect on the surface.
Using the same strategy applied to filled condensates~\cite{Pethick2008}, we calculate the energy of a $\ell=1$ vortex-antivortex pair on the poles of the shell ($\alpha=0$) by integrating the energy of each thin section of height $dz$ along the $z$ axis. We obtain the  energy cost of the vortex as a ratio to the no-vortex state as
\begin{eqnarray}\label{eq:vortex_energy_thick_shell}
E^{\mathrm{shell}}_{\mathrm{v}}/E^{\mathrm{shell}}_0 \approx \frac{2\pi\hbar^2}{3mU_0}\delta \Big( \ln{\frac{R}{\xi_0}} +\ln{\frac{\delta}{\xi_0}} +4.597 \Big),
\end{eqnarray}
where $\xi_0$ is the condensate coherence length, and we assume $\xi_0 \ll \delta$. In the thin-shell limit $\delta\ll R$, the leading term of the vortex energy cost is $\propto \delta \ln R$. Compared with the energy cost $\propto R \ln R$ in the filled-sphere case, the energy cost for a vortex in the thin-shell geometry is much lower than that for a filled condensate with a similar size.

Finally, we study the effects of rotation on 3D shell condensates. We adopt a local-density-approximation method by decomposing a 3D shell into layers of concentric 2D shells. Assuming the vortex string along only the radial direction, the energy cost of a vortex-antivortex pair can be calculated by adding that of each 2D shell section evaluated by Eq.~(\ref{eq:vortex_energy_rotation}). Our results show that a rotating 3D shell condensate still has a critical rotation speed $\Omega_c$ above which the vortex-antivortex pair is stabilized on the poles. Moreover, this critical speed increases with the thickness of the shell. It provides a nondestructive means to measure the thickness of a BEC shell in experiments by finding the lowest rotation speed that stabilizes a vortex-antivortex pair.

In summary, we studied the physics of a vortex-antivortex pair in shell-shaped BECs as the simplest vortex configuration allowed by the closed-surface topology. We found an attractive interaction between the vortex and the antivortex, which causes an energetic instability and hence a self-annihilation of the pair. Such a tendency of self-annihilation can be prevented in a rotating system that has an angular speed above a critical value, where the vortex-antivortex pair is stabilized along the rotating axis. For a 3D shell condensate having finite thickness, we found that the energy cost of producing vortices scales with the thickness and is small compared with that of a filled condensate. Moreover, the critical rotational speed for stabilizing the vortex-antivortex pair depends on the shell thickness, providing an effective means to measure the thickness of shell-shaped condensates in experiments.

\section{Thermodynamics}\label{sec:Thermodynamics}

\begin{figure}[htbp]
  \centering
  \includegraphics[width=\linewidth]{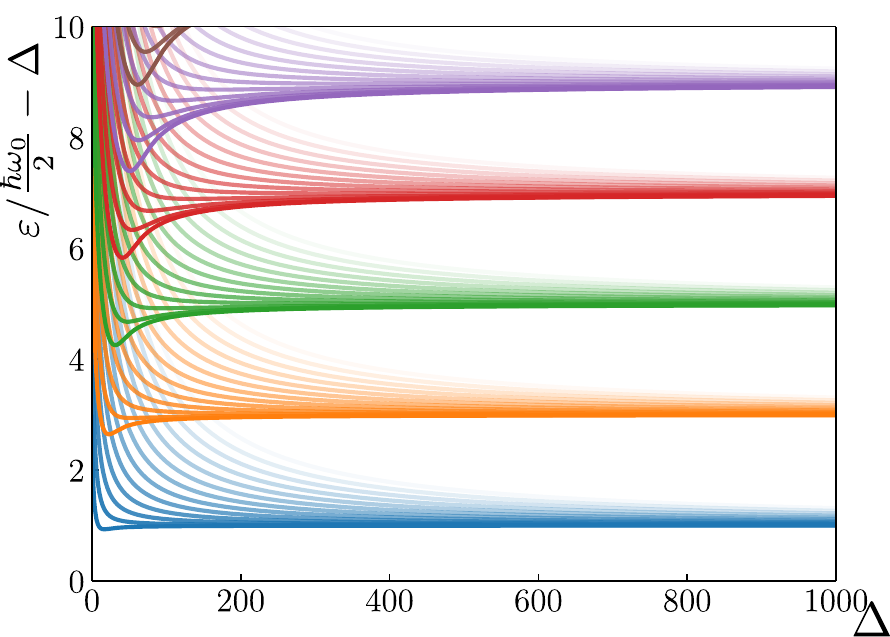}
  \caption{
    (Color online)
    Single-particle energy spectrum, $\varepsilon_{\nu l}$, for non-interacting atoms trapped inside the bubble-trap \cref{eq:bubble}. Here the spectra is plotted against $\Delta$ with $\Omega_b=\Delta$ and the reference potential energy subtracted off. The colors correspond to different values of the radial quantum number $\nu$. For a given $\nu$, we also plot the spectrum for increasing angular momentum $l$ and gradually fade the color in each radial band for clarity.
    In the thin-shell limit, one can see the level-spacing between adjacent radial bands corresponds to that of a simple harmonic oscillator.
  }
  \label{fig:single_particle_spectrum}
\end{figure}

We now turn to thermodynamic aspects of ultracold gases in bubble geometries, their condensate properties, and their evolution in tuning from the filled sphere to the thin shell limits.
For a sufficiently dilute gas, the thermodynamics can be described by neglecting interactions in the Hamiltonian, \cref{eq:Hamiltonian}. In this case, one only needs to solve the single-particle Schr{\"o}dinger equation. As with the collective mode treatment, for a spherically symmetric potential, $V(r)$, the eigenfunctions admit a decomposition of the form
$D_{\nu l}(r) Y_l^{m_l}(\theta,\phi)$,
where the radial component obeys
\begin{align}
    \label{eq:one_body_Schrodinger_eq}
    \varepsilon_{\nu l} D_{\nu l}
    =
    \left( -\frac{\hbar^2}{2m} \frac{1}{r} \partial_r^2 r + V(r) + \frac{\hbar^2 l(l+1)}{2m r^2} \right) D_{\nu l}
    .
\end{align}
For each angular momentum value, we organize the radial quantum numbers such that $\varepsilon_{0,l}\le\varepsilon_{1,l}\le\cdots$.
Over a full range of bubble geometries, we numerically compute the single-particle spectrum in the bubble-trap, \cref{eq:bubble};  the results are shown in \cref{fig:single_particle_spectrum}.
For large detuning frequencies, the bubble trap potential is well approximated by a radially shifted harmonic potential having an effective frequency $\omega_\text{shell} = \omega_0 \sqrt{\Delta/\Omega_b}$.
For asymptotically large $\Delta$, we therefore expect a uniform level spacing structure between adjacent radial bands: $\varepsilon_{\nu+1,l} - \varepsilon_{\nu,l} \sim \hbar \omega_\text{shell}$. For the case $\Omega_b = \Delta $, this simplifies to $\varepsilon_{\nu+1,l} - \varepsilon_{\nu,l} \sim \hbar \omega_0$ at large detunings, which is clearly seen in \cref{fig:single_particle_spectrum}.

From the single-particle spectrum, global thermodynamic quantities can be readily computed at a given temperature $T$ and a chemical potential $\mu$. For instance, the particle number $N$ and entropy $S$ of the system follow from the single-particle spectrum as
\begin{subequations}
    \label{eq:NI_thermodynamics}
    \begin{align}
        N &= \sum_{\nu l} (2l+1) f_{\nu l}
        ,
        \\
        S &= k_\mathrm{B} \sum_{\nu l} (2l+1) \left[ (1+f_{\nu l})\ln(1+f_{\nu l}) - f_{\nu l} \ln f_{\nu l} \right]
        ,
    \end{align}
\end{subequations}
where $f_{\nu l}(T,\mu)=[e^{(\varepsilon_{\nu l}-\mu)/k_\mathrm{B} T}-1]^{-1}$ is the Bose-Einstein distribution function.

\begin{figure}[htbp]
  \centering
  \includegraphics[width=\linewidth]{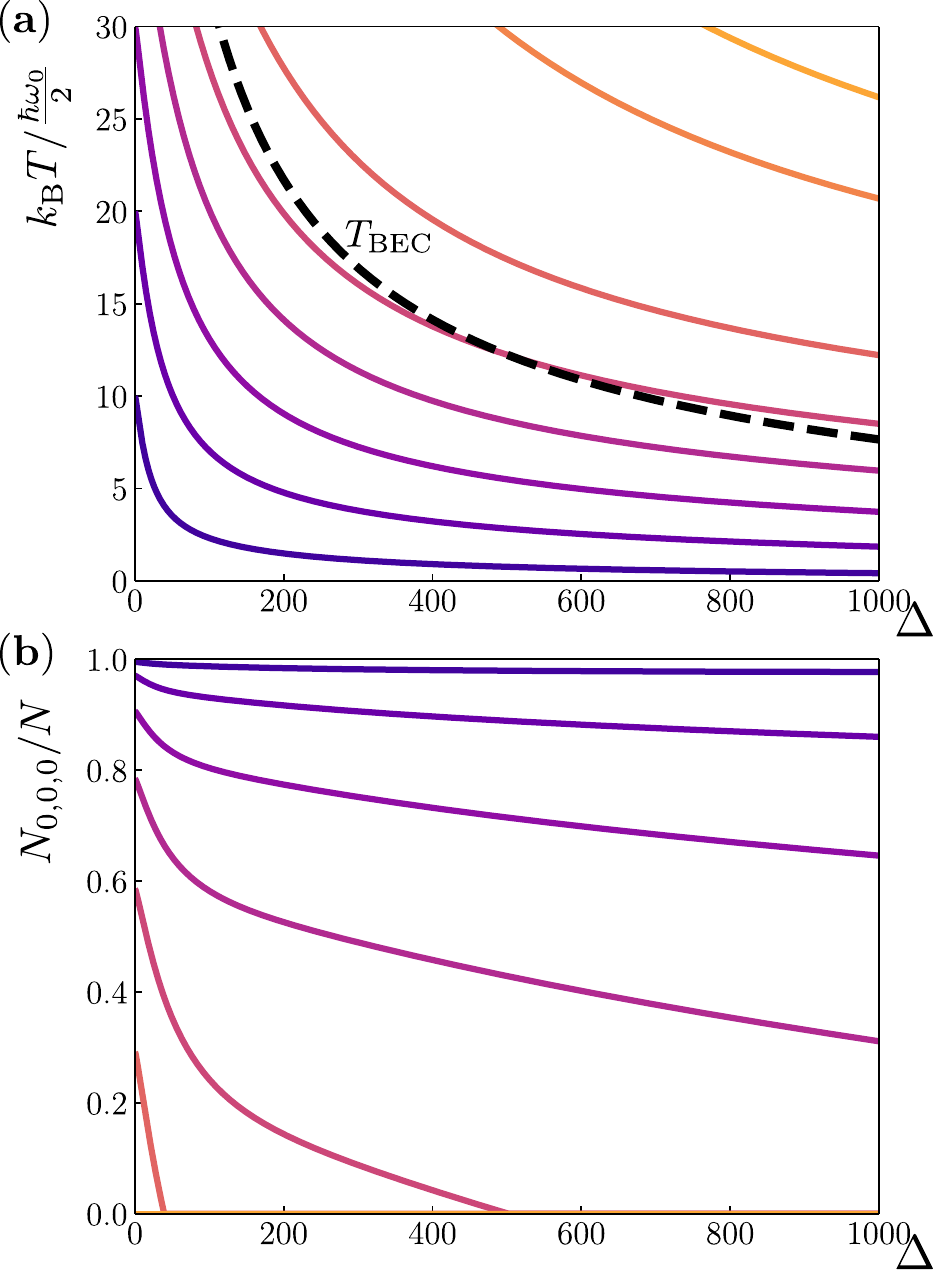}
  \caption{
    (Color online)
    Thermodynamics for $N=5\times10^4$ non-interacting atoms trapped inside the bubble-trap \cref{eq:bubble}.
    (a)
    The BEC critical temperature (black dashed line) is plotted against the detuning frequency $\Delta$ with $\Omega_b = \Delta$.
    In addition, isentropic expansions having different initial temperatures ($k_\mathrm{B} T_i / \frac{\hbar\omega_0}{2} = 10,20,30,\dots$) are shown with colored lines.
    The curves show that although the system cools upon adiabatically expanding the gas into a hollow bubble, the BEC critical temperature drops faster.
    This effect leads to condensate depletion as presented in (b) where the condensate fraction is plotted for the different isentropic expansions considered in (a) using the same color scheme.
    Here $N_{0,0,0}$ refers to the number of atoms in the $(v,l,m_l)=(0,0,0)$ single-particle ground state.
  }
  \label{fig:noninteracting_thermodynamics}
\end{figure}

In \cref{fig:noninteracting_thermodynamics}, we present the thermometry of the system as a function of bubble size. First, we compute the BEC critical temperature, determined by fixing the particle number and solving the implicit equation
\begin{align}
    N = \sum_{\nu l \neq 0,0} (2l + 1) \frac{1}{e^{(\varepsilon_{\nu l} - \varepsilon_{0,0})/k_\mathrm{B} T_\mathrm{BEC}}-1}
    ,
\end{align}
where the single-particle ground state is excluded from the sum.
As can be discerned from \cref{fig:noninteracting_thermodynamics}, the critical temperature decreases with increasing bubble size.
This is consistent with the evolution from a 3D ball to a quasi-2D thin bubble and an associated increase in low-energy density of states (see the spectra in \cref{fig:single_particle_spectrum}).
The trend can also be argued simply on density grounds. The volume of the condensate can be estimated at large detunings using the approximate radially shifted oscillator form of the potential. One finds that the volume scales as $\Delta^{3/4} \Omega_b^{1/4}$~\cite{Rhyno2021}, or simply $\Delta$ in our case. Hence, for fixed atom number, the density $\sim N / \Delta$ drops during the expansion process, lowering $T_\mathrm{BEC}$.

Next, we consider adiabatic processes by fixing both the particle number and entropy of the system as the bubble geometry changes from a thick-to-thin shell.
Fixing the entropy makes such expansions isentropic and hence adiabatic.
To carry out this calculation, we fix the particle number and consider various initial  temperatures for the system , starting at the filled sphere limit $\Delta = 0$.
When the initial system is Bose-condensed, we compute the entropy and condensate fraction, and when it is in the normal phase, we compute the entropy and chemical potential.
Having derived the entropy, we can modify the trap parameters and determine the new temperature and condensate fraction ($T<T_\mathrm{BEC}$) or temperature and chemical potential ($T>T_\mathrm{BEC}$) for an isentropic process by solving \cref{eq:NI_thermodynamics} for fixed $N$ and $S$.
Importantly, for an initially Bose-condensed gas, we observe that isentropic expansions exhibit condensate depletion~\cite{Rhyno2021}.
In \cref{fig:noninteracting_thermodynamics} we see this effect in the $\Omega_b = \Delta$ case.
In particular, the temperature of the system decreases during isentropic expansions, as does the BEC critical temperature.
However, $T_\mathrm{BEC}$ drops faster than $T_\mathrm{isentropic}$ and hence this leads to a loss of phase-space-density.
This is manifested clearly in the behavior of the condensate fraction during adiabatic bubble inflation, where Bose-condensed systems initially closer to the transition (i.e. smaller initial condensate fraction) cross more quickly into the normal phase upon expansion.
Although these considerations neglect interactions, in the thin-shell regime at fixed particle number, a more careful analysis using approximate solutions to the Bogoliubov equations leads to the same conclusions~\cite{Rhyno2021}.

In summary, we discussed the thermodynamic properties of a dilute bubble-trapped Bose gas.
Employing solutions to the Schr\"odinger equation, we were able to compute various thermodynamic quantities relevant to addressing the experimental feasibility of Bose-condensed bubbles, such as the critical temperature and condensate fraction.
Consistent with our previous work, we showed that isentropic bubble inflation leads to condensate depletion, and hence to a loss of phase-space density.

\section{Nonequilibrium dynamics}\label{sec:Nonequilibrium_dynamics}

In studying the evolution from filled to thin shells, we now go beyond equilibrium thermodynamics. In our equilibrium treatment of the previous section, we computed observables in the grand canonical ensemble using the density operator
$\RR = \frac{1}{\mathcal{Z}}e^{-(\HH-\mu \hat{N})/k_\mathrm{B}T}$, where 
$\mathcal{Z} = \tr e^{-(\HH-\mu \hat{N})/k_\mathrm{B}T}$
is the grand partition function and the Hamiltonian is assumed static. 
Here we consider a system initially in thermal equilibrium at temperature $T$
and chemical potential $\mu$
whose Hamiltonian then changes dynamically in time.

Given the wider application of the nonequilibrium method developed for this section and its multiple non-trivial aspects, we provide a semi-pedagogical presentation below. 
In particular, we compute observables subject to the density operator $\RR(t) = \UU(t,t_0) \, \RR(t_0) \, \UU^\dag(t,t_0)$, where
$\RR(t_0)$
is the initial thermal state at time $t_0$, and
$\UU(t,t_0) = \mathcal{T} \exp(-\frac{i}{\hbar} \int_{t_0}^{t} dt' \HH(t'))$ is the time evolution operator with $\mathcal{T}$ denoting time-ordering.

Such density operators necessarily describe isentropic processes as the entropy remains constant under unitary time evolution: $S(t) = - k_\mathrm{B} \tr(\RR(t) \ln \RR(t)) = S(t_0)$.
Although the entropy is fixed, it does not mean that all predictions of this nonequilibrium theory match those of the previous section.
This is because, upon reaching some Hamiltonian $\HH(t)$ at a later time, the dynamic density operator $\RR(t)$ generally will not take on the same thermal form.
Furthermore, the dynamics considered here will not necessarily be adiabatic in the quantum mechanical sense in which a system initialized in an eigenstate tracks the instantaneous eigenstate under sufficiently slow changes in the Hamiltonian.

For a sufficiently dilute atomic Bose gas in the bubble trap, \cref{eq:bubble}, we model the system with the Hamiltonian
\begin{align}
    \hat H(t) = \int d\mathbf{r}
    \,
    \hat \psi^\dagger(\mathbf{r}) \left( -\frac{\hbar^2}{2m}\nabla^2 + V_\text{bubble}(r,t) \right) \hat \psi(\mathbf{r})
    ,
\end{align}
where
$V_\text{bubble}(r,t)$ is the bubble trap potential \cref{eq:bubble} with a time-dependent detuning frequency $\Delta(t)$ (again, we take $\Omega_b = \Delta$ throughout).
Although we concentrate here on the case of a time-dependent spherical bubble trap, much of the formalism that follows can be extended to arbitrary time-dependent trapping potentials.

At all times $t$, i.e. all detunings $\Delta(t)$, the instantaneous single-particle eigenstates are characterized by the quantum numbers $\nu , l, m_l$, and the Hamiltonian can be diagonalized as
\begin{align}
    \HH(t) = \sum_{\nu l m_l} \varepsilon_{\nu l}(t)
    \,
    \bb^{S \dag}_{\nu l m_l}(t) \bb^S_{\nu l m_l}(t)
    ,
\end{align}
where we define bosonic ladder operators in the Schr\"odinger picture as
\begin{align}
    \bb^S_{\nu l m_l}(t) \equiv
    \int d\mathbf{r}
    \,
    [D_{\nu l}(r,t) Y_l^{m_l}(\theta,\phi)]^*
    \,
    \hat\psi(\mathbf{r})
    \,
    .
\end{align}
Here $D_{\nu l}(r,t)$ denote instantaneous solutions to \cref{eq:one_body_Schrodinger_eq}.
Note that the instantaneous radial component of the single-particle eigenfunctions, $D_{\nu l}(r,t)$, which are computed numerically, are only uniquely determined up to an overall phase. Thus, caution is required to ensure that these eigenfunctions (or the bosonic ladder operators constructed from them) are not treated as differentiable with respect to time.

Here we focus on computing the instantaneous occupation of various $\nu,l,m_l$ modes,
\begin{align}
    N_{\nu l m_l}(t)
    &= \tr \left[ \RR(t_0) \, \bb^{H \dag}_{\nu l m_l}(t) \bb^H_{\nu l m_l}(t) \right]
    \label{eq:Heisenberg_picture_mode_expectation_value}
    ,
\end{align}
where we introduce Heisenberg operators in the usual way: $\bb^H_{\nu l m_l}(t) \equiv \UU^\dag(t,t_0) \, \bb^S_{\nu l m_l}(t) \, \UU(t,t_0)$.

To accomplish this task, it will be convenient to introduce another set of Heisenberg operators defined as
\begin{align}
    \label{eq:Heisenberg_operators_for_initial_eigenbosons}
    \bb^H_{\nu l m_l}(t,t_0) \equiv 
    \UU^\dag(t,t_0) \, \bb^S_{\nu l m_l}(t_0) \, \UU(t,t_0)
    ,
\end{align}
which time-evolve the bosonic ladder operators that diagonalize the \textit{initial} Hamiltonian to the present time.
Importantly, the Heisenberg equations of motion for the operators in \cref{eq:Heisenberg_operators_for_initial_eigenbosons} do not require differentiating the Schr\"odinger ladder operators and the arbitrary overall phase in the eigenfunctions causes no issues.
The Heisenberg equations of motion can be formally solved, with the result being
\begin{align}
    \label{eq:Heisenberg_initial_eigen-bosons}
    \bb^H_{\nu l m_l}(t,t_0) = \sum_{\nu'}
    \bigl[ \mathcal{U}_{l}(t,t_0)
    \bigr]_{\nu\nu'}\,
    \bb^S_{\nu' l m_l}(t_0)
    .
\end{align}
The unitary matrix in the above equation is given by the time-ordered exponential
\begin{align}
    \label{eq:TOE_initial_eigen-bosons}
    \mathcal{U}_{l}(t,t_0)
    &\equiv \mathcal{T} \exp\left( -\frac{i}{\hbar} \int_{t_0}^t dt' \, \mathcal{H}_l(t',t_0) \right)
    ,
\end{align}
where
$\mathcal{H}_l(t,t_0) \equiv \mathcal{O}_l^\dag(t,t_0) \, \mathcal{E}_l(t) \, \mathcal{O}_l(t,t_0)$
is a Hermitian matrix with
$\mathcal{E}_l(t)$
being diagonal and composed of instantaneous eigenvalues of \cref{eq:one_body_Schrodinger_eq}
and $\mathcal{O}_{l}(t,t_0)$ a unitary composed of overlap integrals between the instantaneous and initial eigenfunctions of \cref{eq:one_body_Schrodinger_eq}.
Specifically, the matrix elements are given by
$ [\mathcal{E}_{l}(t)]_{\nu\nu'}
\equiv \delta_{\nu\nu'} \, \varepsilon_{\nu l}(t)$
and
$ [\mathcal{O}_{l}(t,t_0)]_{\nu\nu'}
\equiv \int_0^\infty dr \, r^2 \, D^*_{\nu l}(r,t) D_{\nu' l}(r,t_0)$.
By expanding the bosonic ladder operators that instantaneously diagonalize the Hamiltonian in terms of those that diagonalize the initial Hamiltonian one finds that the Heisenberg operators in \cref{eq:Heisenberg_picture_mode_expectation_value} take the form
\begin{align}
    \label{eq:Heisenberg_instantaneous_eigen-bosons}
    \bb^H_{\nu l m_l}(t) = \sum_{\nu'}
    \bigl[
    \mathcal{O}_{l}(t,t_0)
    \,
    \mathcal{U}_{l}(t,t_0)
    \bigr]_{\nu\nu'}\,
    \bb^S_{\nu' l m_l}(t_0)
    .
\end{align}

Equipped with \cref{eq:Heisenberg_instantaneous_eigen-bosons} it is straightforward to compute the dynamics of n-point correlation functions, at least formally.
Simply expand each Heisenberg operator in terms of the ladder operators that diagonalize the initial Hamiltonian using \cref{eq:Heisenberg_instantaneous_eigen-bosons} and then compute the resulting n-point correlators for the initial thermal state using e.g. Wick's theorem.
Focusing on the instantaneous occupation of the $\nu,l,m_l$ modes, \cref{eq:Heisenberg_picture_mode_expectation_value}, we find $N_{\nu l m_l}(t)$ is given by
\begin{align}
    \left[
    \mathcal{O}_{l}(t,t_0)
    \mathcal{U}_{l}(t,t_0)
    \,\, \mathrm{diag}(f_{\nu l}(t_0)) \,\, 
    \mathcal{U}^\dag_{l}(t,t_0)
    \mathcal{O}^\dag_{l}(t,t_0)
    \right]_{\nu\nu}
    ,
\end{align}
where $f_{\nu l}(t_0)$ is the Bose-Einstein distribution function evaluated for the initial thermal state.

\textit{Drive protocol}: to demonstrate the method, we choose to drive the system out of equilibrium by considering a discrete linear quench protocol:
$\Delta(t) = s \delta\Delta$, if $t_{s-1} < t \le t_s$, where $t_s = s \delta t$ with $s=1,2,\dots,P$ (setting $t_0 = 0$).
For such discrete protocols, the time-ordered exponential in \cref{eq:TOE_initial_eigen-bosons} decomposes into a product of matrix exponentials:
\begin{align}
    \label{eq:TOE_initial_eigen-bosons_drive_protocol}
    \mathcal{U}_{l}(t_s,t_0)
    &=
    e^{ -\frac{i}{\hbar} \delta t \, \mathcal{H}_l(t_s,t_0)
    }
    \cdots
    e^{ -\frac{i}{\hbar} \delta t \, \mathcal{H}_l(t_1,t_0)
    }
    .
\end{align}
We also consider $\omega_0 \delta t \delta \Delta < 1$, so that we can contextualize the physics of the nonequilibrium dynamics approximately in terms of the analogous continuous linear quench protocol with $\Delta(t) = v_Q t$ with quench rate $v_Q = \delta\Delta / \delta t$.

\begin{figure}[htbp]
  \centering
  \includegraphics[width=\linewidth]{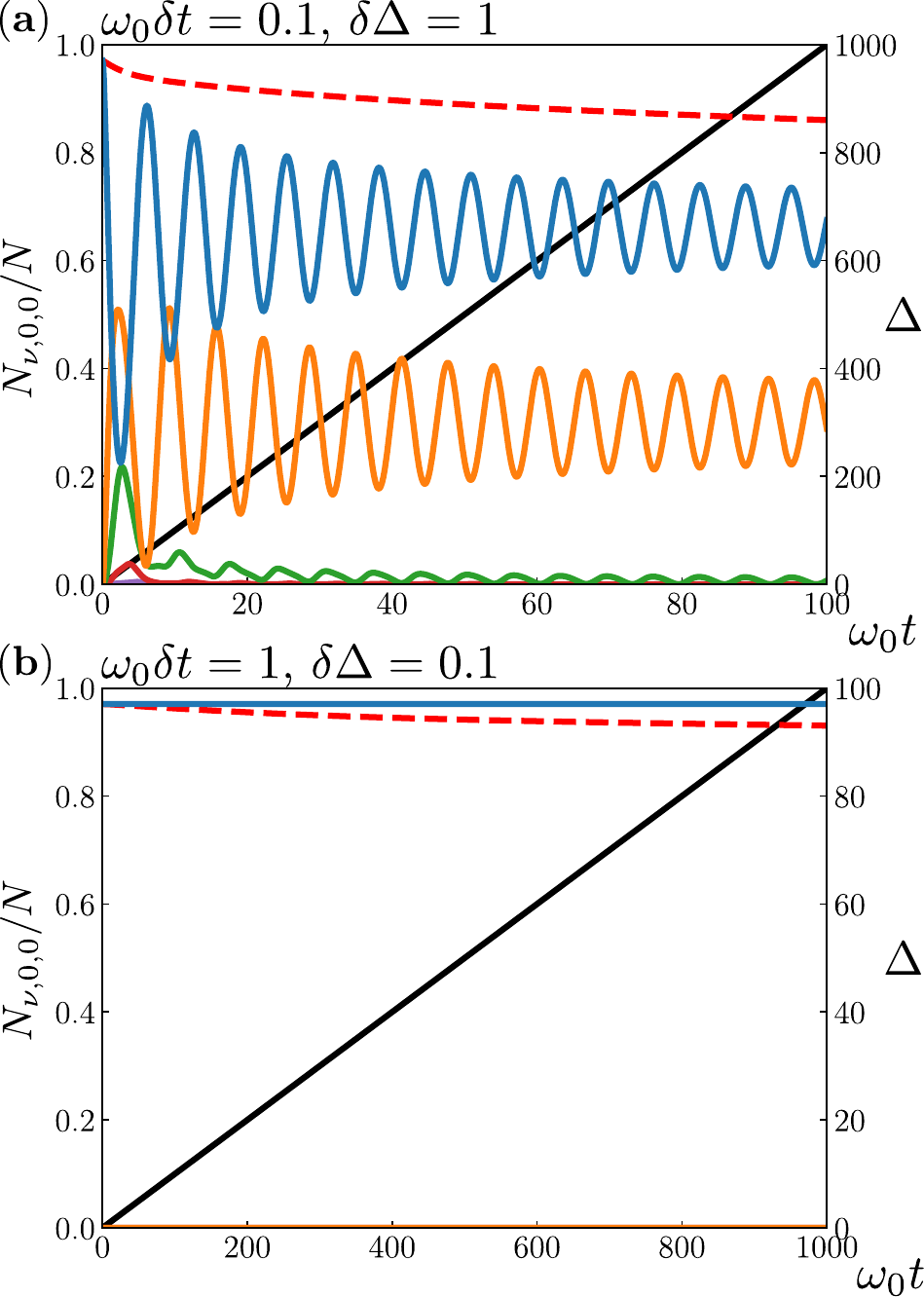}
  \caption{
    (Color online)
    Nonequilibrium quench dynamics for atoms trapped inside a time-varying bubble-trap. Here we consider an initial thermal state with $N=5\times10^4$ atoms at temperature $k_\mathrm{B} T = 20 \frac{\hbar\omega_0}{2}$.
    (a) and (b) show two different quenches with parameters $\omega_0 \delta t = 0.1$, $\delta\Delta=1$ and $\omega_0 \delta t = 1$, $\delta\Delta=0.1$ respectively.
    The detuning parameter $\Delta(t)$ is shown with a black solid line and marked on the right axis.
    The instantaneous occupation fraction for various radial modes with $l=m_l=0$, $N_{\nu,0,0}(t)/N$, are shown with colored lines. Here the color scheme for different radial quantum numbers $\nu$ matches that of \cref{fig:single_particle_spectrum}.
    (a) The quench rate, $v_Q = \delta \Delta / \delta t = 10 \omega_0$, is large compared to the radial gap which leads to nonequilibrium excitations of higher-energy modes.
    (b) The quench rate, $v_Q = 0.1 \omega_0$, is small compared to the radial gap which leads to a nearly constant condensate fraction throughout the evolution.
    For comparison, in both (a) and (b) the isentropic thermodynamic condensate fraction, found by solving \cref{eq:NI_thermodynamics}, is shown with a red dashed line.
  }
  \label{fig:quench_dynamics}
\end{figure}

In \cref{fig:quench_dynamics}, starting from an initial thermal state with $N=5\times10^4$ atoms at temperature $k_\mathrm{B} T = 20 \frac{\hbar\omega_0}{2}$ ($\approx 97\%$ initial condensate fraction), we display the dynamics of $N_{\nu l m_l}(t) / N$ for the $l=0$ angular momentum state using two separate discrete linear quenches.
In the first protocol, \cref{fig:quench_dynamics}(a), we take $\omega_0 \delta t = 0.1$, $\delta\Delta=1$, which corresponds to a quench rate of $v_Q = 10 \omega_0$. Whereas, in the second quench protocol, \cref{fig:quench_dynamics}(b), we take $\omega_0 \delta t = 1$, $\delta\Delta=0.1$, which corresponds to a quench rate of $v_Q = 0.1 \omega_0$.
[In both cases, we consider $P=1000$ steps, which is sufficiently large that the piecewise nature of each function is not shown in the figure.]
In the first case, \cref{fig:quench_dynamics}(a), where the quench rate is large compared to the radial gap in the single-particle spectrum (see \cref{fig:single_particle_spectrum}), the instantaneous condensate fraction drops below that of isentropic thermodynamic predictions, and higher-energy radial modes are excited.
One also observes oscillations as atoms transition between populating the ground state and excited states.
In contrast, in the second case depicted in \cref{fig:quench_dynamics}(b), where the quench rate is small compared to the radial gap, higher-energy states are not appreciably excited throughout the evolution and the condensate fraction remains nearly constant.
This marks a mixed-state realization of the quantum adiabatic theorem.
Compared with the results of the previous section, where the condensate depletes during isentropic thermodynamic expansions, this nonequilibrium quench results in a condensate fraction that is higher than its equilibrium counterpart.

In summary, we have outlined a method for computing nonequilibrium correlation functions for noninteracting bosons in a bubble trap with an initial thermal density operator.
We confined ourselves to bubble trap geometries, but the method can be straightforwardly generalized to arbitrary traps.
We also considered only a single type of discrete quench protocol.
In future work, one can envision employing discrete quench protocols that cannot be interpreted through a continuous counterpart,
taking interactions into account and studying collective effects of dynamically expanding and contracting BEC bubbles,
studying observables as the system is quenched across the BEC phase transition and the ensuing physics of the Kibble-Zurek mechanism~\cite{Kibble1976,Zurek1985,Dziarmaga2010,Polkovnikov2011,Zurek2014},
engineering shortcut to adiabaticity protocols~\cite{GueryOdelin2019},
and so much more.

\section{Experimental realizations and outlook}\label{sec:Experimental_realizations}

Finally, having surveyed our theoretical explorations of the rich range of phenomena hosted by shell-shaped BECs, we turn to some current experimental settings for realizing these structures. Each system comes equipped with capabilities for probing one set of phenomena over another, be they exciting collective modes via impulse or vortices either spontaneously or through rotation, or observing thermodynamics or non-equilibrium dynamics via tuning trapping parameters, or many other possibilities. Combining theory work presented here as well as other extensive work, the community is well-poised for experiment and theory to work together in comprehensively unearthing the range of phenomena unique to BEC shells. 

\begin{figure}[htbp]
  \centering
  \includegraphics[width=\linewidth]{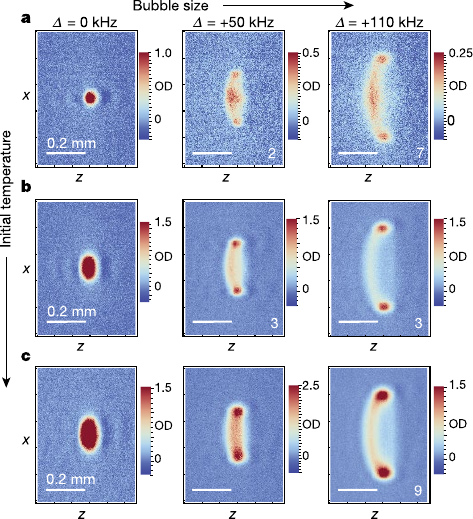}
  \caption{
    (Color online)
    Ultracold atomic bubbles experimentally observed aboard the ISS using CAL.
    Each image shows optical depth measurements in the $x$-$z$ plane with subsequent rows corresponding to larger initial temperatures for the bubble inflation process (a: $\sim \SI{100}{\nano K}$, b: $\sim\SI{300}{\nano K}$, c: $\sim\SI{400}{\nano K}$).
    Subsequent columns correspond to larger detuning frequencies with shell structures emerging at the largest detunings.
    (Reused from Ref.~\cite{Carollo2022}. Copyright (2022) by Springer Nature.)
  }
  \label{fig:CAL_bubbles}
\end{figure}

Realization of ultracold bubble structures has primarily been driven via the technique of rf-dressing of magnetically-trapped atomic ensembles, originally proposed in 2001~\cite{Zobay2001,Zobay2004}. This technique has yielded gravitationally-sagged shell structures in terrestrial labs at multiple points in previous decades~\cite{DeMarco2006,Colombe2004,Merloti2013,Harte2018,Beregi.2024}, and is a primary tool in ongoing work with CAL aboard the ISS~\cite{Carollo2022}. The rf-dressing technique is well-described in multiple reviews~\cite{Perrin2017,Garraway2016}, and typically involves tailored chips of MHz-scale radiofrequency of polarization non-collinear with the local DC magnetic field (in the case of CAL, from an atom chip trapping potential).
The CAL results of 2022, as reproduced in Fig.~\ref{fig:CAL_bubbles}, depict bubbles possessing significant asymmetry and inhomogeneity~\cite{Carollo2022}.
The aspect-ratio inhomogeneity is largely driven by the architecture of the atom-chip magnet wiring, and the residual inhomogeneity is largely driven by spatial variation of the rf field amplitude~\cite{Lundblad2019}. Current experiments aboard CAL are focusing on the observed nonadiabaticity of the bubble inflation (dressing) process.
Coming upgrades to hardware aboard CAL will feature improved shell aspect ratio, improved rf field homogeneity, and improved imaging~\cite{PrivateCommunication}.

Orbital microgravity experiments benefit from the atoms experiencing a perpetual free-fall environment, but experimental modifications to devices such as CAL are costly and time intensive.
A promising alternative approach is to achieve temporary free-fall conditions in terrestrial drop towers such as the Einstein-Elevator at the Hannover Institute of Technology~\cite{Lotz2017,Lotz2018,Lotz2020,Lotz2023}.
This facility is a multi-user \SI{40}{\meter} tall drop tower capable of performing up to 300 tests per day under microgravity conditions.
At present, researchers are actively working to utilize the Einstein-Elevator as an alternative route for rf-dressed BEC bubble production.

Recent experimental advances reported by Wang’s group have created and explored shell-shaped BECs in a two-species atomic gas~\cite{Jia2022,Huang2025}.
In their work, mixtures of $^{23}$Na and $^{87}$Rb atoms were prepared in a spherical optical trap, where repulsive interspecies interactions caused an immiscible state to form resulting in a shell-shaped $^{23}$Na BEC with a $^{87}$Rb BEC core.
They employed a magic-wavelength optical dipole trap to minimize the displacement between the centers of mass of each BEC, hence mitigating the adverse effects of gravitational sag and producing a spherical geometry on Earth.
In a recent experiment~\cite{Huang2025}, they probed the hollowing transition of a shell-shaped $^{23}$Na BEC using collective excitations.
By properly modulating the trapping potential and the interspecies scattering length, they successfully induced in-phase and out-of-phase collective modes of the $^{23}$Na BEC shell.
They observed a dip in the out-of-phase oscillation frequency spectrum, which serves as a robust signature for the onset of the hollowing transition, the point where the $^{23}$Na BEC transitions from a filled sphere to a hollow shell surrounding the $^{87}$Rb core.
In this setting, the $^{87}$Rb core can be treated as an effective potential experienced by the $^{23}$Na shell.
In such a case, the experimental findings agree with the physics of single-species shell-shaped BECs discussed in Sec.~\ref{sec:Collective modes}, confirming a universal dip structure in the frequency spectrum at the hollowing transition regardless of potential details.

In addition to the experimental testbeds described above, it would be worth revisiting some of the earliest experimental efforts, which hinted at the possibility of realizing shell-shaped BECs in optical lattices. In such a system, ultracold bosonic atoms are loaded in deep optical lattices, where the interplay among the tunneling energy, interatomic interaction, and chemical potential results in a BEC state or a Mott-insulator state of the atoms~\cite{Sachdev2001}. With a spherical trap turned on, the two states are expected to coexist, forming a thin BEC shell embedded in a Mott-insulator structure~\cite{Batrouni2002,DeMarco2005,Campbell2006,Barankov2007,Sun2009}. This prediction, together with the spectroscopic evidence of the condensate order in the system, has sparked theoretical investigations on the physics of shell-shaped BECs. Nevertheless, the unique features of a shell-shaped BEC in optical lattices, such as those in its equilibrium profile and collective excitations, remain unexplored in experiments.

Coming full circle with shell-shaped BECs in space, a fascinating prospect entails stellar bodies. Here, exceedingly high densities that show a steep gradient across the body from its core to the exterior not only allow for concentric shells of different phases of matter, while the stellar temperatures are incredibly high compared to terrestrial ones, they are still low enough to support condensates.
Neutron stars, for instance, are predicted to host shells of superfluid as well as superconducting matter~\cite{Sauls1989,Pethick2017}.
Glitches in rotation frequencies of associated pulsar bodies have been attributed to the reconfiguration of tangles of quantized vortices in shell-shaped condensate regions~\cite{Alpar1984}.
Theoretical work on shell-shaped BECs and their experimental realizations in controlled settings offer prototypes for studying speculated behavior within stellar bodies.
Furthermore, in mimicking cosmic phenomena such as the predicted structure formation via Kibble-Zurek physics or particle pair production in inflationary models, they offer a playground for studying the Universe itself~\cite{Visser2002,Schmiedmayer2013,Eckel2018,Feng2018,Banik2022,Viermann2022}.

\begin{acknowledgments}

We thank Giuseppe De Tomasi and Matteo Sbroscia for illuminating discussions.
This work was supported by the National Aeronautics and Space Administration (NASA) Science Mission Directorate, Division of Biological and Physical Sciences (BPS) through ROSES-22 as well as multiple Jet Propulsion Laboratory (JPL) Research Support Agreements (B.R., J.B., N.L., and S.V.).
B.R. and N.G. gratefully acknowledge funding from the DLR Space Administration with funds provided by the Federal Ministry for Economic Affairs and Climate Action (BMWK) under grant numbers 50WM2451 (QUANTUMANIA) and from the Deutsche Forschungsgemeinschaft (DFG) through SFB 1227 (DQ-mat) within Project A05, Germany’s Excellence Strategy (EXC-2123 QuantumFrontiers Grants No. 390837967).

\end{acknowledgments}

\vspace{0.5 cm}

\section*{Author Declarations}

\subsection*{Conflict of interest}

The authors have no conflicts to disclose.

\section*{Data Availability}

The data that support the findings of this study are available from the corresponding author upon reasonable request.

\bibliography{references}

\end{document}